\documentclass[openacc]{rstransa}

\usepackage{graphicx,amsmath,relsize,epstopdf,color,mathtools,newtxtext,newtxmath,braket,rotating}

\usepackage{siunitx}
\usepackage[nolist]{acronym}
\usepackage{multirow}
\usepackage{dcolumn}
\usepackage{hyperref}
\usepackage{color}

\titlehead{Research}

\begin{document}

\title{An operational distinction between quantum entanglement and classical non-separability}
\author{Natalia Korolkova$^{1,3}$, Luis~S\'anchez-Soto$^{2,3}$, Gerd Leuchs$^{3,4,5}$}
\address{$^{1}$School of Physics and Astronomy, University of St. Andrews, North Haugh, St. Andrews, KY16 9SS, UK\\
$^{2}$Departamento de \'Optica, Facultad de F\'{\i}sica, Universidad Complutense, 28040 Madrid, Spain\\
$^{3}${Max-Planck-Institut f\"{u}r die Physik des Lichts, 91058~Erlangen, Germany}\\
$^{4}${Institut f\"ur Optik, Information und Photonik, Universit\"at Erlangen-N\"{u}rnberg, 91058 Erlangen,  Germany}\\
$^{5}$Department of Physics, University of Ottawa, Ottawa, {Ontario K1N 6N5}, Canada}
\subject{Quantum Physics, Quantum Optics, Vector Optics}
\corres{Natalia Korolkova\\
\email{nvk@st-andrews.ac.uk}}
\begin{abstract}
Quantum entanglement describes superposition states in multi-dimensional systems – at least two partite – which cannot be factorized and are thus non-separable.  Non-separable states exist also in {classical theories} involving vector spaces. In both cases, it is possible to violate a Bell-like inequality. This {has} led to controversial discussions, which we resolve by identifying an operational distinction between the classical and quantum {cases}. 
\end{abstract}


\begin{fmtext}

\section{Introduction}
In 1997, Spreeuw first {discussed}  ``a classical analogy of entanglement''~\cite{Spreeuw1997}, followed four years later by a paper elaborating on the classical analogy of quantum information processing using wave optics~\cite{Spreeuw2001}. {Depriving quantum entanglement} of its purely quantum nature and {exploring the new capabilities} of classical optics {have sparked a} range of intriguing theoretical considerations and experiments (for reviews, see,  e.~g., \cite{ProgressOpt,AielloNJP2015,wir2019,Shen2022}).

{This exploration has prompted a controversial discussion}, the essence of which we highlight with a few selected examples. In 2011, Qian and Eberly suggested a new interpretation of light 
\end{fmtext}
\maketitle
\noindent 
polarization based on ``non-quantum entanglement," stating that  ``polarization is a characterization of the correlation between the vector nature and the statistical nature of the light field."  {Subsequent elaborations} on the emerging links between quantum and classical optics~\cite{Eberly2015,Eberly2016} {led} Eberly et al. to a strong statement: ``... entanglement is a vector space property, present in any theory with a vector space framework. There is no distinction between quantum and classical entanglements, as such"~\cite{Eberly2016}. This assertion has {elicited} equally strong {opposing statements}. {As has been highlighted in discussions on this topic in Science~\cite{Karimi2015}, “Entanglement is a property of the quantum world; classical systems need not apply". Karimi and Boyd wrote, referring to "classical entanglement" term: “We do not endorse this new nomenclature," adding, “... this situation lacks the key feature - nonlocality"~\cite{Karimi2015}.} 

Despite the {wide range of opinions, the past decade has witnessed} a plethora of potentially useful experiments with vector optics and other classical implementations of vector spaces, demonstrating the power of 'classical entanglement' (or rather non-separability) in high-precision measurements, metrology applications, and in emulation of various quantum information protocols. However, the controversy in {interpreting} these experiments and discussions about the fundamental nature of entanglement, {both} quantum and classical, {remains} ongoing~\cite{Shen2022}.

We {address} the controversy by recognizing that the term {should rely solely on the details of the experimental observation rather than any prior knowledge about the system.} As we will explain, concepts like non-locality or counting the number of particles involved {fail to offer} a clear distinction between quantum and classical entanglement.

Instead, there is a decisive difference: the characterization of quantum entanglement always involves  correlating statistical outcomes of multiple measurements on a given system, as {elaborated} below. {Conversely,} classical entanglement typically refers to deterministic correlations between a single measurement and a filter or sorter operation, such as those imposed on a light beam by a polarizing beam splitter or another quantum sorter~\cite{Berkhout2010,Ionicioiu2016}. The {settings of such filters or sorters} do not involve measurement (and is a unitary operation). This provides a clear distinction between the two phenomena. 

 \section{Entanglement revisited: questions to ask} 
 \label{ent-revisited}

The term entanglement {refers to a unique form of correlation between} two or more different variables (for an excellent recent discussion of quantum entanglement, see \cite{Paneru_2020}). Mathematically, entangled states are described by non-separable functions within the vector space known as Hilbert space. Their global nature {leads to the puzzling} property that the outcome of a measurement on one part of such a non-separable state at one location {appears to have an instantaneous} effect on the other part of the state at another location, resulting in perfect correlations between measurement results at both locations. However, this "immediate action" cannot be detected solely by measuring at one location.

{The emergence of the `non-local effect' from a local measurement stems} from quantum theory, {giving rise} to several conceptual questions. {Opponents}  of this {difficult-to-grasp} {action at a distance} brought up explanations {rooted in classical} physics. This was used by John Bell  to demonstrate that quantum physics cannot be explained by classical stochastic theories relying on local hidden variables, hence {coining} the term quantum entanglement~\cite{bell_aspect_2004}. 

The non-local {nature} of correlations in quantum entanglement stems from non-separability. However, similar types of correlations can be found in a classical description involving mathematically non-separable functions. Are we {discussing here} correlations of a common nature, as suggested, for example, by {Eberly}? The term 'classical entanglement' was coined. Is such terminology appropriate? Does any mathematical non-separability imply Bell-type non-locality, and is it this non-locality that makes up entanglement?

\begin{figure}[t]
\centering
\includegraphics[width=0.8\linewidth]{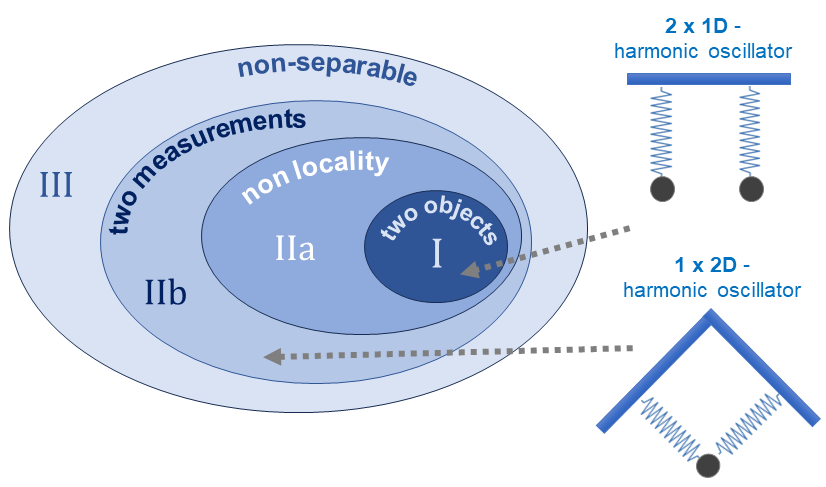}
\caption{Left: Different possible subsets of a set of general non-separable states. The ellipses represent the four decisive properties of systems under consideration: consist of two objects; exhibit non-locality; allows for two measurements; can be written as mathematically non-separable mode functions. The examples of physical systems populating a particular set are described in detail in the text:
I - non-separable states involving two excitations (two objects); IIa,b - non-separable states with only one single excitation (object), but where two distinct measurements can still be performed on two non-separable partitions of the Hilbert space. III - classically entangled states described by non-separable mode functions. Right: Two paradigmatic examples of the states of type I and IIb, both allowing for two distinct independent measurements - see text for discussion.
\label{mengen}}
\end{figure}

John Bell emphasized {that his analysis applies to situations involving} two distinctly different particles. This {establishes} a clear {and} naturally unique partitioning of the Hilbert space of the composite quantum system, as well as the {measurements to be performed.} However, {the choice of partitioning} becomes less obvious when different degrees of freedom contribute to the dimensions of the Hilbert space, particularly in discussions involving light and mode functions. {Extending these discussions to such scenarios has recently gained attention, partly in relation to the debate on quantum versus classical entanglement.}

In a recent study by Paneru et al.~\cite{Paneru_2020}, quantum contextuality has been used {as a tool} to identify {the} genuinely quantum nature of correlations between different degrees of freedom. Realism, locality, and non-contextuality for single and multiple particles {represent} different aspects of physical theories. {Paneru et al.} argue that classical states of light, {whether} separable or non-separable, {can} be fully described using the wave picture {without} invoking field quantization. {Therefore, they} cannot challenge these fundamental concepts in classical physics (for an introduction to field quantization and the quantum theory of light,  see~\cite{Loudon}).

We want to address the problem from a different, more hands-on viewpoint. Realism, locality, and non-contextuality are all related to a measurement process. Figure~{\ref{mengen}} visually highlights some of the nontrivial aspects of partitioning and the role of measurement in analyzing correlations in non-separable systems. For simplicity, we restrict ourselves to bi-partite non-separable states. Different possible partitions of the {\emph{mathematical}} vector space can be of quite different {\emph{physical}} {nature~\cite{Goldberg:2022aa}}. They can correspond to objects of zero or non-zero rest mass, such as a photon or an atom, respectively, or to mode functions, such as particular spatial modes. 

{The crucial role of partitioning  has first been explicitly highlighted early 2000s as ambiguity and hence freedom in selecting multipartite structure for a given system.\footnote{Not so much thinking about different degrees of freedom and their nature.} This renders also the presence of entanglement and its degree ambiguous \cite{Zanardi1, Zanardi2}.  In \cite{Zanardi1} Zanardi writes: ``Clearly even the notion of entanglement is affected by
some ambiguity being relative to the selected multipartite structure...The above ambiguity is removed as soon as, according to some criterion, a preferred multipartite structure is
selected among the family of all possible partitions into subsystems.''  These papers hence aimed at presenting a way to control the amount of achievable entanglement via the partitioning choice. Their main body is devoted to develop a corresponding mathematical set-up, based on the ideas of virtual quantum systems and compoundness notion \cite{Zanardi1}, leading to a general algebraic framework where the desired quantum tensor product structure is {\it observable induced} \cite{Zanardi2}. The interested reader is referred to  \cite{Zanardi1, Zanardi2}, as the presented theory is beyond the scope of our work and not relevant for our discussion here. We share however one of the key realisation, stated already in these papers: ``a partitioning of a given Hilbert space is induced by the experimentally accessible observables (interactions and measurements)''\cite{Zanardi2}. Or putting it differently: ``the system $S$ is viewed as composed  by $S_1$, $S_2$, . . . if one has some operational access (is able to
“access,” “control,” “measure”) to the individual degrees of freedom of $S_1$, $S_2$''\cite{Zanardi1}.} 

{For the task we set here, operational distinction between quantum and classical non-separability,  this crucial point should be formulated in a different way:  it is not the partitioning choice but the physical nature of the given subsystems that is a key issue. This consideration becomes particularly profound in the realm of optics.} To re-iterate, the complication stems from the following observation: non-separability is a mathematical property and is independent of the physical nature of the different degrees of freedom; i.e., the different parts of Hilbert space, whereas the ability to perform one or more measurements does depend on the physical nature of the degree of freedom. A measurement requires some excitation, some energy above the level of the vacuum, which can be associated either with a zero or non-zero rest mass particle or a combination thereof. Mode functions in optics{, for example,} can have numerous degrees of freedom occupying a larger Hilbert space, but they are merely frames for excitations and by themselves they do not refer to any excitation. If the mode is not {excited,} there is nothing specific to measure (other than non-specific vacuum fluctuation). {We, therefore,} attempt to structure the types of non-separable states with regard to their physical nature. We take {the liberty here} to use the word `excitation" in a wider sense. We speak about excitation not only when we have some non-zero energy state of the electromagnetic field, but we can also view a massive particle as a de Broglie wave-packet. We can then also think of ``modes" which are ``excited" or ``not excited," meaning the particle is there or not. Furthermore, there is a whole range of quasi-particles, meaning a combined excitation from entities of different physical natures: phonons, excitons, polaritons, to name a few. {To emphasize this aspect we are writing the wave functions also in the following form, underlining that the degrees of freedoms provide the space where excitations can live:}

\begin{equation}
\begin{aligned}
\vert \Psi\rangle & = \vert \rm excitation \rangle_{\rm degrees~of~freedom} 
\label{where excitations live}
\end{aligned}
\end{equation}

We list below the properties defining principally different cases of non-separability and refer to excitations to accommodate massive particles, photons or ``inter-species'' excitations.

\subsection{Defining properties of non-separable systems of different types}

We want to single out four defining properties {that lead} to non-separable states of principally different physical nature (see Figure~{\ref{mengen}}). We {will} elucidate their distinctive features with examples in the next subsection. 

\paragraph{Two objects.}
The dark-blue innermost set in {Fig.~{\ref{mengen}} encompasses} systems with two distinct objects {(i.e., excitations)} living in the sub-spaces of a bi-partitioned Hilbert space. {These objects} are in a state described by mathematically non-separable mode functions. Here, the available `modes'  are populated by two {excitations,} rendering 
two distinct separate measurements (each on one of the excitations) possible. Examples of such system are two {electrons in a singlet state} (spin up/down, $\vert \uparrow \rangle, \vert \downarrow \rangle $) or two photons of orthogonal {polarization} (polarization {horizontal/vertical}, $\vert H \rangle, \vert V \rangle $) in different spatio-temporal modes. {In the literature we usually see them in the following form:}
\begin{equation}
\begin{aligned}
\vert \Psi_{\rm singlet} \rangle_{AB} & = \frac{1}{\sqrt{2}} \left ( \vert \uparrow \rangle_A  \vert \downarrow \rangle_B  -  \vert \downarrow \rangle_A  \vert \uparrow \rangle_B \right ),  \\ 
\vert \Psi_{\rm pol} \rangle_{AB} & = \frac{1}{\sqrt{2}} \left ( \vert H \rangle_A  \vert V \rangle_B  -  \vert V \rangle_A  \vert H \rangle_B \right ).
\label{singlet}
\end{aligned}
\end{equation}
{In order to facilitate the comparison between different scenarios, we will use the notation of Eq.(\ref{where excitations live}) so that the above reads in terms of `modes' and excitations as:} 
\begin{equation}
\begin{aligned}
\vert \Psi_{\rm singlet} \rangle & = \frac{1}{\sqrt{2}} \left ( 
\vert 1 \rangle_{\rm A,\uparrow}
\vert 1 \rangle_{\rm B,\downarrow}
-  
\vert 1 \rangle_{\rm A,\downarrow}
\vert 1 \rangle_{\rm B,\uparrow}
\right ), \\
\vert \Psi_{\rm pol} \rangle & = \frac{1}{\sqrt{2}} \left ( 
\vert 1 \rangle_{\rm modeA,H}
\vert 1 \rangle_{\rm modeB,V}
-  
\vert 1 \rangle_{\rm modeA,V}
\vert 1 \rangle_{\rm modeB,H}
\right ).
\label{singlet-modes}
\end{aligned}
\end{equation}
{Here, for example, the mode index \{$A,\uparrow$\} denotes `a mode' understood as a framework for excitation at location $A$ in the form of a particle in spin up (see Eq.~\eqref{where excitations live} and discussion before it). In the corresponding Dirac ket, $\vert 1 \rangle_{\rm A,\uparrow}$, ``$1$'' means that this mode is indeed filled with this excitation, that is, particle in spin up state is present at location A. The ket $\vert 0 \rangle_{\rm A,\uparrow}$ would mean no spin-up particle at $A$. In the second equation, `a mode' \{$\rm modeA,H$\} is a spatial mode $A$ framing an excitation corresponding to a photon in $A$, polarized horizonatally, with $1$ in the ket meaning ``this mode is excited'', that is, we have a horizontally polarized photon.}

The mathematically non-separable mode functions in Eq.~(\ref{singlet-modes}) correspond to the possible {combinations} of outcomes of two projective measurement defined by the excitations present. This subset, depicted in Fig.~\ref{mengen} by the dark central {ellipse}, marks the subset {that} we historically perceive as {\it THE} entanglement. Elements of this subset {possess all} the four properties mentioned in {Fig.~\ref{mengen}:} (1) {they consist} of two distinctly different objects; (2) two measurements are possible; i.e., each of the two subsystems into which the system is partitioned can be measured separately; (3) {they are} non-local; (4) {they are} non-separable. Correlations  are manifested through series of two measurements performed on the two subsystems involved and constitute {\it statistical} correlations. 

\paragraph{Non-local.} 
The next medium-blue shaded subset {includes} all states that exhibit the property of non-locality (in its trivial sense: parts of a composite system are at spatially separated locations). {This} {may require merely} {one photon \cite{Bjork}:}
\begin{equation}
\begin{aligned}
\vert \Psi_{\rm pol} \rangle & = \frac{1}{\sqrt{2}} \left ( 
\vert 1 \rangle_{\rm modeA}
\vert 0 \rangle_{\rm modeB}
-  
\vert 0 \rangle_{\rm modeA}
\vert 1 \rangle_{\rm modeB}
\right ).
\label{non-local}
\end{aligned}
\end{equation}

We briefly discuss non-locality in more general sense in {Sec}.~\ref{nonloc}, but it is deliberately not the subject of this paper. 

\paragraph{Two measurements.} The larger {ellipse} of the yet lighter blue includes all states characterized by a common {feature: the} possibility to perform two distinct, independent projective measurements. Remarkably, two of the three subsets contained within the {ellipse} labelled `two measurements' deprive {us of} the first traditional ingredient: two objects. Here, in these two lighter-shaded subsets, there is only one single excitation initially. However, importantly, it still clearly defines two Hilbert space partitions. These partitions are determined by two distinct excitations generated by one single object, examples {are given in the next section (see second line in Eq.~\eqref{peres})}. The mathematically non-separable mode functions  describe the possible combination of measurement outcomes defined by these secondary excitations.

\paragraph{Non-separable.} For states in the outermost {ellipse} (but outside the other subsets), we now {relinquish} the possibility {of performing} two measurements. In this {class,} the only {remaining} property is mathematical non-separability, which is also common to all cases listed above.{ This non-separability} 
{is not tied} to {excitations, and may even be demonstrated with a single coherent state excitation $\vert \alpha \rangle$, for example as} { discussed in \cite{wir2019} at the end of sec. 4.1.: decomposition in a particular basis followed by sorting/filtering determined by this basis can yield correlations emulating those of entangled state:}
\begin{equation}
\begin{aligned}
\vert \Psi_{\rm non-sep} \rangle & = 
\vert \alpha \rangle_{\left ( {\rm modeA,H} \right ) - \left ( {\rm modeB,V} \right )}.
\label{non-separable}
\end{aligned}
\end{equation}
{Here, the mode function characterized by spatio-temporal mode A and horizontal polarization is written as $\{ {\rm modeA,H }\} $. } {For details see \cite{wir2019}.}

 \subsection{Paradigmatic examples {of} non-separable systems} 
 \label{examples}
 

In this {subsection, we aim to explore} specific physical implementations of the non-separable states encompassed in different  subsets in {Fig.}~\ref{mengen}, {emphasizing the features that define} their entanglement properties.

\paragraph{Subset I.} 
This is an example of {prototypical} quantum entanglement, {the most familiar among} physical implementations of non-separable states. {These states belong} to the type {depicted in Fig.~\ref{mengen}, top left,} and presented in {Eq.~\eqref{singlet}.}

\paragraph{Subset IIa.} 
Consider a single excitation shared by two mode functions, involving  non-separability. The seminal example of such quantum entanglement is {the} entanglement of a single photon ({or} atom) incident on a beam splitter. The state at the output of the beam splitter can be written as $\vert \psi_{12} \rangle \propto \vert 1 \rangle_1 \vert 0 \rangle_2 + \vert 0 \rangle_1 \vert 1 \rangle_2$ { (see Eq.~\eqref{non-local})}, where $\vert 1 \rangle_j$ refers to a photon (or atom) in output $j$ and $\vert 0 \rangle_j$ represents no photon (or atom), with $j=1,2$.

In optics, there has been a debate {on whether this scenario truly} represents quantum entanglement, a controversy that was resolved in 2004~\cite{Bjork}. This {debate is relevant to our current discussion}, as the argument against entanglement {was based on} only one excitation being present and measurable at the output with on-off single photon detectors. However, as it is well-known,  an optical beam splitter always has two inputs and two outputs. The incident light field excitation {determines} the set of modes we are looking at and, consequently, the two output modes of the beam splitter.

In {the} case of an `undivisible' single photon, one output mode {contains} a single photon {while the other exhibits vacuum} fluctuations in the mode defined by this excitation and vice versa. Both {of these secondary} excitations (with one corresponding to the ground state) can be measured using {homodyne detection}. {Thus,} although only one physical excitation is present {(i.e., one photon),} {two measurements can still} be performed at the output using homodyning, {providing evidence} of entanglement (see also \cite{getting-used}). This example IIa, {also features} the property of non-locality (see {Fig.~\ref{mengen})}.

\paragraph{Subset IIb.}  
In this {scenario,} we consider a single massive object {that undergoes} further excitations in two ways; e.g., motion in two dimensions. A seminal example {of this is} a single {two-dimensional} harmonic oscillator with possible excitations in both the $x$- and $y$-directions. {An intriguing} observation made by Asher Peres in his textbook~\cite{Peres1995} nearly 30 years ago (and seldom cited) {clarifies} this example and  {underscores} the significance of two measurements defined by two excitations {in discussions} about quantum entanglement. The right {part} of Fig.~\ref{mengen} displays the two systems Peres discusses. Two {one-dimensional (1D)} harmonic oscillators coupled in a non-separable way (Fig.~{\ref{mengen}}, top right) is a generic example of quantum entanglement, ticking all four boxes as in dark blue subset. However, mathematically this system is completely equivalent to a single mass on two springs, a single {two-dimensional (2D)} oscillator (Fig.~{\ref{mengen}}, bottom right) \cite{Peres1995}. Consequently, the wave functions of the energy {eigenstates} are completely equivalent in {both} cases:
\begin{equation}
\begin{aligned}
\vert \Psi_{\rm 2 \times 1D} \rangle_{AB} & = \frac{1}{\sqrt{2}} \left ( \vert n \rangle_A  \vert m \rangle_B  -  \vert m\rangle_A  \vert n \rangle_B \right ),   \\ 
\vert \Psi_{\rm 1 \times 2D} \rangle_{xy} & = \frac{1}{\sqrt{2}} \left ( \vert n \rangle_x  \vert m \rangle_y  -  \vert m\rangle_x  \vert n \rangle_y \right ).
\label{peres} 
\end{aligned}
\end{equation}
Here, $\{\vert n,m \rangle\}$ denotes two possible energy levels associated with two excitations, either in the two oscillators {at locations $A$ and $B$}, for two 1D systems, or with two spatial directions $x$ and $y$ of a 2D system.

Crucially, {similar to example IIa,} there are two distinct secondary excitations. Hence, {despite containing only one mass, this system} allows for two distinct measurements to be performed, thus decisively manifesting
the entanglement property.\footnote{In this example we view the quantum harmonic oscillator as a discrete-variable (DV) system {due to its discrete energy levels. Alternatively, one can view it as a continuous-variable (CV) system by looking at} position and momentum, which requires a different framework. Note that a light mode in quantum optics can be represented as such a CV quantum harmonic oscillator~\cite{Loudon}, and in example IIa, the two output modes can be viewed as such two coupled harmonic oscillators.}

{A 2D oscillator} provides a prototypic example of such states, but the subset is not restricted to 2D systems. Non-separable states involving quantum entanglement between two different degrees of freedom exhibit the same property. A vivid example of a {locally} entangled state is a mesoscopic Schr\"odinger cat-like state of cold atoms realized with trapped Be ions in 1996 in the group of Wineland~\cite{wineland}. They created the following state of a single $^9{\rm Be}^+$ laser-cooled ion:
\begin{equation}
\vert \psi \rangle= \frac{1}{\sqrt{2}} ( \vert x_1 \rangle \vert \uparrow \rangle + \vert x_2 \rangle \vert \downarrow \rangle  )  \, .
\label{atom}
\end{equation}
{Here}, $\vert x_{j} \rangle$ {denotes} localized wave-packet states corresponding to two spatial positions of the atom, while $ \vert \uparrow \rangle,  \vert \downarrow \rangle$ {are} two distinct internal electronic quantum states of the atom (hyperfine ground states). The secondary excitations present for this single atom can be modelled as {a} 2D harmonic oscillator or even interpreted as a {Schr\"odinger cat-like state} of two spatially separated coherent harmonic oscillator states, $\vert x_{j} \rangle$, $j=1,2$ (as suggested in~\cite{wineland}). The key point here is that two measurements on {separate exitations of} two different degrees of freedom, $\vert x_{j} \rangle$ and $ \vert \uparrow \rangle,  \vert \downarrow \rangle$,  are possible for the state \eqref{atom}. {Therefore, the state \eqref{atom} is} prototypic for ``local entanglement'' of states which cannot be separated spatially. {In other words, } such states {not only lack the} ``two-objects'' property, but {also lack another} traditional feature, non-locality.

\paragraph{Subset III.}  
All {the examples discussed are characterized} by mathematically non-separable functions, {as shown in Fig.~\ref{mengen}. Now, consider} the region of the full set outside the three inner subsets. {In contrast to the previous examples,} here the partition of the Hilbert space is not defined by two excitations, and is thus {representing a case} of classical entanglement.

For classically entangled non-separable states, {the} Hilbert space is spanned by the {available mode functions}, and a single excitation, which can, in principle, be assigned to (included in) any of the partitions. {This assignment} is guided solely by experimental convenience or our choice of which measurements to perform.

Let us illustrate this with an example. In a number of metrology tasks, { performance enhancement} has been demonstrated using a classically entangled state with correlations between the transverse spatial mode and polarization. {This is the situation formally displayed in Eq.~\eqref{non-separable}. Often, however, the distinction between excitation and mode functions is not carved out so explicitely and one finds the following notation (see, e.g., \cite{AielloNJP2015})}:
\begin{equation}
\vert \Psi \rangle \propto \vert \psi_{01}( \mathbf{r}) \rangle_A \vert H \rangle_B - \vert \psi_{10}(\mathbf{r}) \rangle_A \vert V \rangle_B.
\label{cl-ent1}
\end{equation}
As we {mentioned at the} beginning of our discussion {on Fig.~\ref{mengen}, quantum mechanics stems from observations, namely measurements.} In the case of Eq.~\eqref{cl-ent1}, we {are dealing} with the non-separability of a vector space described entirely by available mathematical mode functions connected to different degrees of freedom. { {The excitation} enabling a measurement and the non-separable mode functions as a whole form a product state. This excitation, thus, does not contribute to the entanglement}.

Contrast {this with all} the previous cases: {there, the} partitioning of the Hilbert space has been based on physical excitations. The Dirac kets corresponded to the chosen bases for the respective vector sub-spaces, and their eigenvalues were the possible measurement outcomes of the two projective measurements, one on each partition.

For  Eq.~\eqref{cl-ent1}, the partitioning initially is not connected to any excitations and the {eigenvalues} of the basis kets are  {\it mode-functions, possible decomposition of the initial state in some basis}. As we will discuss in detail later, one of the eigenvalues is {\it the conditional measurement outcome of one single projective measurement on a chosen partition},  conditioned on the deterministic choice of the value of the other variable. That is, to unlock the physical nature of the {state Eq.~\eqref{cl-ent1} and to connect it to observations, we rather need to view this equation in the spirit of Eq.~\eqref{non-separable}}.  

{The} Hilbert space has a partition 1 spanned by two mode functions $k$ and $k^\prime$ corresponding to the first degree of freedom (spatial mode), and partition 2 spanned by two mode functions $l$ and $l^\prime$ corresponding to the second degree of freedom (polarization mode). {However, the excitation} present is independent of these partitions and corresponds to the electromagnetic wave, which can be written in the basis of various modes. Hence, the physics behind Eq.~\eqref{cl-ent1} is better captured {by rewriting it as}
\begin{eqnarray}
\vert \Psi \rangle \propto 
[ \vert \psi_k \rangle_1 \vert \psi_l \rangle_2   - \vert \psi_{k^\prime} \rangle_1 \vert \psi_{l^\prime} \rangle_2  ] \; \vert { E(\mathbf{r}, t)} \rangle,
\label{cl-ent2}
\end{eqnarray}
where $E(\mathbf{r}, t)$ describes the excitation of the electromagnetic field. Put simply, we have decomposed  the electromagnetic wave into two spatial modes: $\psi_{01}$ {horizontally} polarized and $\psi_{10}$ vertically polarized {(cf Eq.~\eqref{non-separable})}.  {Modifying the notation in this way brings us a bit closer to Eq.~\eqref{where excitations live}, which clearly emphasizes what is the mode (subscript) and what is the excitation (argument of the ket).} In this system, we cannot violate realism,  a cornerstone of classical physics absent in quantum mechanics, {Here, we are dealing with measurement outcomes predetermined before measurement, evident in the correlations observed.}

{This is the key difference to the quantum entanglement of Eq.~\eqref{atom}, which is {also local but which displays two correlated excitations. Classical non-separability of Eqs.~\eqref{non-separable},~\eqref{cl-ent1},~\eqref{cl-ent2} is associated with only this one excitation}. Exactly this difference is captured in Figures~\ref{cl-ent} and \ref{q-ent} - we need to compare the measurement procedures, as we elaborate in detail in the next section. The two states are deceptively similar:  Both of them do not posses the non-locality property and exhibit correlations between two internal degrees of freedom. Both can be can written as a non-separable state in a composite vector space. However, for the quantum state of Eq.~\eqref{atom}, the partitioning of Hilbert space is dictated by {more than one excitation present, hence two independent measurements can be performed, one on each partition, and their statistics and correlation can be verified (see also \cite{Zanardi1,Zanardi2} and our discussion on it in Sec. 2). As we already highlighted above, the situation is drastically different in the case of classical entanglement of Eq.~\eqref{cl-ent1}; and this is not a question of being able to perform a certain type of measurement or not, but rather a problem of having nothing specific to measure in one of the partitions.}

{Note that the latter statement is also relevant to differentiate between IIa and III. For IIa, the input beam splitter modes determine a vacuum mode at {one of the outputs of this beam splitter alongside a photon in the other one and vice versa. Tracing out one of the output modes, the remaining output mode will be in a mixed state excited by a fraction of a photon on average. Both these correlated output modes, each with a fractional excitation, can be measured demonstrating the correlation, provided the right measurement is done, which is homodyne detection in this case \cite{Bjork}. In a way, in case of states in III, there is a free choice of where to assign the one and only available excitation $\vert { E(\mathbf{r}, t)} \rangle$, as illustrated by Eq.~\eqref{cl-ent2}. Formally, the excitation can be assigned to any one of the  mode function} partitions. Let us elaborate on what this implies for a measurement process.}

To elucidate the relationship between the spatial mode and polarization, 
we can opt for one of the two possible scenarios. First, assume we {aim} to measure polarization. We {assign} excitation to partition 2 and {conduct measurement} $\hat{M}_2$ in the basis $\lbrace H, V \rbrace $, where potential outcomes are denoted ${\cal M}^j_2$, with $j=H,V$. However, {to discern} the correlation between spatial mode and polarization, {one} must first choose a specific spatial mode to examine: either $\psi_{01}(\mathbf{r})$ or $\psi_{10}(\mathbf{r})$ from partition 1. This is not a measurement but {rather} a unitary operation~\footnote{{A sorter or filter introduces a certain projection in the corresponding vector space, resembling quantum measurement. However, in a sorter, the state is projected and the complementary projection is in principle still available, such as in a polarizing beam splitter in optics. This operation can be reverted if both output beams are used. In a quantum measurement one is necessarily ‘loosing’ information which is in principle there before the measurement. The quantum measurement projects the state of the system onto an eigen state of the operator describing the measurement. Often a sorting or filtering operation is followed by a measurement leading to some kind of read out. The readout 'causes' the collapse or projection of the state of the system to just one possibility. With sorting only, without any readout, all possibilities would still be open. (Note, that if you were to look only at the read out in one of the output ports of a sorter, what you see is independent of what is done at the other sorter outputs. This, by the way, is the reason why entanglement does not allow one to transfer information faster than the speed of light. }} {(in contrast, combining the projections in the Hilbert sub-space with a measurement as in previous examples, the evolution is always non-unitary, as for any }projective measurements).  The projection onto one of the axes in partition 1 determines the axis onto which the projection in partition 2 occurs, thus predetermining the measurement outcome on 2 before any measurement is performed: a unitary operation on 1 selecting $\psi_{01}(\mathbf{r})$ implies ${\cal M}^H_2$.

Similarly, {if our aim is to resolve} the type of transversal mode in the measurement, we will assign excitation to partition 1, {where potential measurement outcomes are denoted} ${\cal M}^j_1$, with $j=01,10$. The projection {onto} one of the axes in partition 2 determines the axis onto which the projection in partition 1 occurs: a unitary operation on 2 selecting $H$ implies ${\cal M}^{01}_1$ prior any measurement. Note {that} only one measurement on one of the partitions is possible in this case. {We will  delve into this case further} in the next two sections.

{Another property to consider, besides non-separability, is}  non-locality, which one would {attribute} to (I) and (IIa), but not to (IIb) and not, in general, to the entire set. {It is worth noting that in cases (I), (IIa), and (IIb), two separate independent measurements can be performed.}   


From the discussion and examples above, {it is evident} that there are inherently different questions {one needs to distinguish} when seeking to assess the nature of correlations:
\begin{enumerate}
\item Quantum-classical demarcation. What is a proper verification tool here?

\item Is entanglement entirely about non-separability of {states, regardless of} the physical character of the mathematical partition of a vector space?

\item Is entanglement about {the} non-local character of correlations? What does `non-local' mean? What are appropriate means to verify it?

\item When employing a non-separable state in certain protocols, where does the advantage {come} from?
\end{enumerate}

\section{{Measurements} to reveal quantumness}

\begin{figure}[t]
\centering
\includegraphics[width=0.8\linewidth]{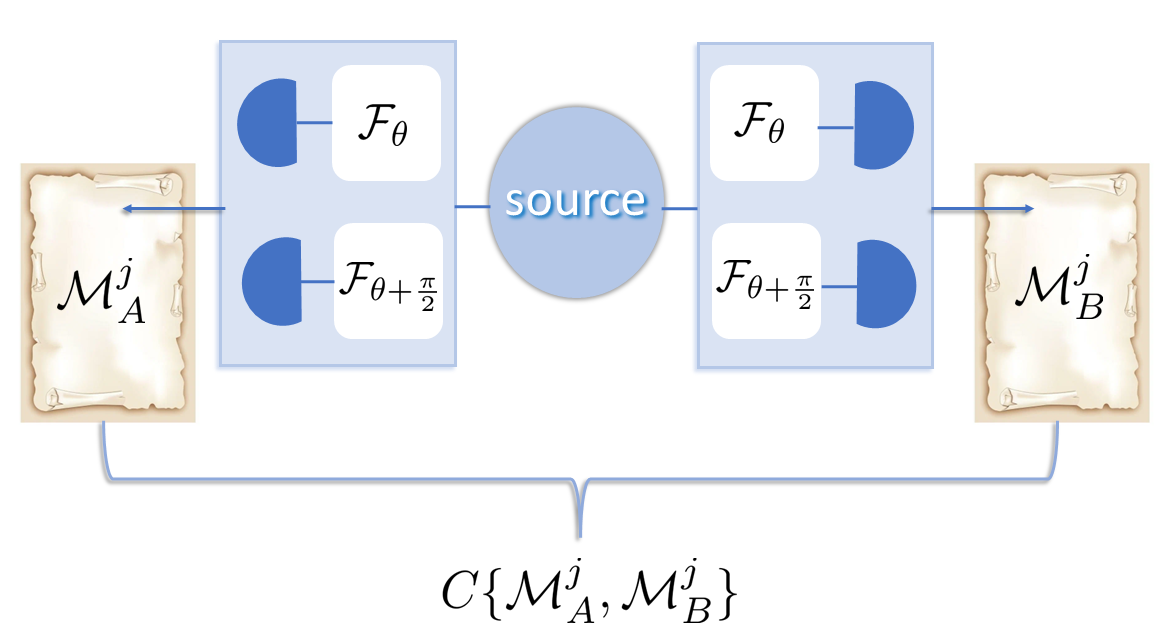}
\caption{Correlations in quantum entanglement. Two distinct projective measurements on subsystems $A,B$ are performed depicted by two boxes, with a possibility of detection in two bases of choice, ${\cal F}_{\theta}, {\cal F}_{\theta + \frac{\pi}{2}} $, and with statistical outcomes  ${\cal M}_{A,B}^j$. There are statistical correlations between measurement outcomes, $C \lbrace {\cal M}_A^j, {\cal M}_B^j \rbrace$.
\label{ent}}
\end{figure}

In this {section, we address the first two previous} points. We {begin} by emphasizing that the Bell inequality {provides answers} to very specific questions (as will be addressed in the next section) and requires extreme caution {when applied to verify the quantumness of correlations}~\cite{Khrennikov}.
{Therefore we take here a measurement-based, operational approach which has common points with earlier work of \cite{Zanardi1,Zanardi2} and recent contributions by \cite{Khrennikov,Khrennikov2}. The approach of \cite{Zanardi1,Zanardi2} is based on observable induced choice of Hilbert space partitioning as already discussed in section 2. The most recent work of Khrennikov and Basieva \cite{Khrennikov2}  brings the idea of observable induced partitioning \cite{Zanardi2} even further and aims to decouple the notion of entanglement from the tensor product structure and associated non-separability. Their work emphasizes the role of measurement process even more: Khrennikov and Basieva develop the probabilistic formalization of the Schr\"odinger's concept  of entanglement without reference to any complex vector spaces. Entanglement is defined as dependence of observables $A$ and $B$; relevant constraints are derived based on the quantum conditional probability calculus. In our approach and the approaches referred above, choice of  observables, operational access to degrees of freedom, and hence a measurement process as such, step forward as pivotal tools to assess quantum entanglement presence.}

\begin{figure}[t]
\centering
\includegraphics[width=0.7\linewidth]{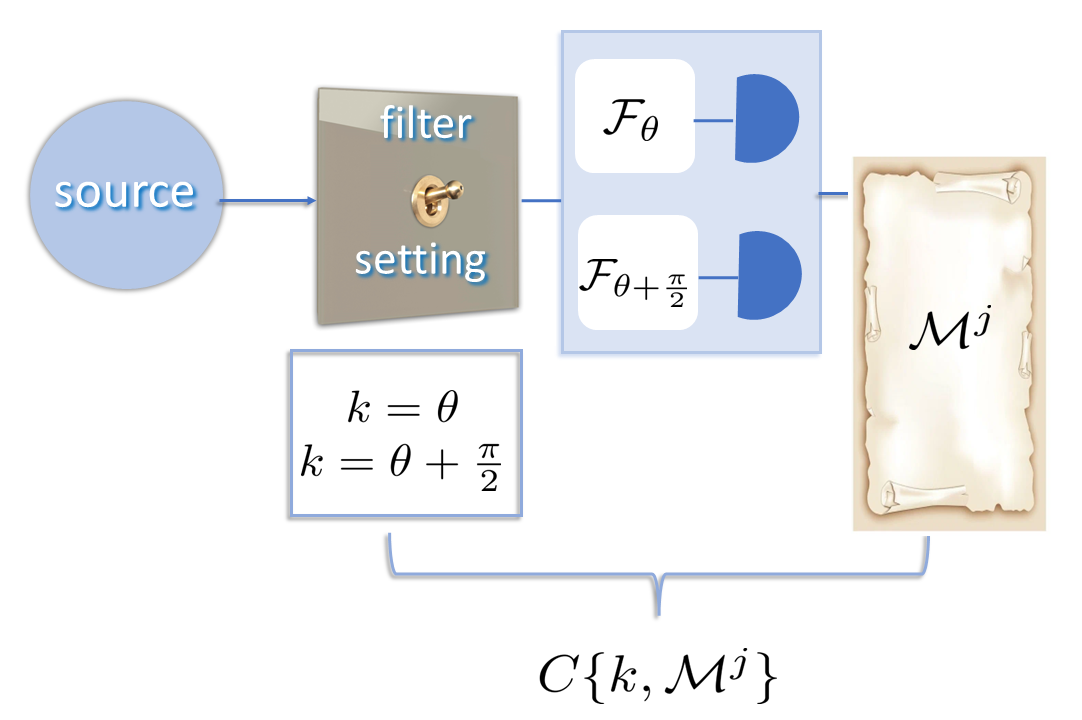}
\caption{Correlations in classical entanglement. Only one projective measurement described by mode function $B$ is performed, denoted by a box with a possibility of detection in two bases of choice, ${\cal F}_{\theta}, {\cal F}_{\theta + \frac{\pi}{2}} $ and with outcome ${\cal M}^j$. The other mode function, $A$, describes a unitary operation of ``a switch'', determining the filter setting $k$, ${\cal F}_{\theta}$ or ${\cal F}_{\theta + \frac{\pi}{2}}$. There is a deterministic correlation between the filter setting $k$ and the  measurement outcome, $C \lbrace k, {\cal M}^j \rbrace$.
\label{cl-ent}}
\end{figure}

{In assessing quantumness,} a system cannot be considered separately from the measurement process. The key {lies in the interaction between} a quantum system and a measurement apparatus, {with quantum mechanics providing predictions for the corresponding outcomes. Additionally, the experimental context, including the details of the apparatus used, is crucial, as it directly influences} the observation process.

To assess a {quantum} property, one has to be sensitive to the projection imposed by a quantum measurement. This cannot be achieved {by a single measurement alone; rather, one} needs to perform a first measurement to impose the projection and then a second measurement to detect the effect of the projection. This is {precisely} what occurs in the case of quantum entanglement. A projective measurement is performed on one part of the non-separable state,  $A$. {Subsequently,} a second projective measurement is performed on the other part, $B$, revealing the effect of the first measurement on the global state. Overall, quantum correlations inherent to entangled states are manifested through the statistically correlated results of subsequent projective measurements. {Mathematically,} in the partitioning of the vector space, both mode-functions $A$ and $B$ in the non-separable state are {eigen states} of projectors, {with their eigenvalues (say, $0$ and $1$)} describing the possible outcomes of the projective measurements, denoted as ${\cal M}^{0,1}_A$ and ${\cal M}^{0,1}_B$, respectively, one on each of the partitions (see {Fig.~{\ref{ent}})}:
\begin{equation}
\vert \Psi \rangle \propto \vert 0 \rangle_A \vert 1 \rangle_B + \vert 1 \rangle_A \vert 0 \rangle_B. \label{q-ent}
\end{equation}
{We see that the} correlations are revealed between  two statistical measurement outcomes, $C \lbrace {\cal M}_A^j, {\cal M}_B^j \rbrace$.

{In the case of classical entanglement,} the vector space is partitioned in a {fundamentally} different manner (as discussed in subsection \ref{examples}). There is only one excitation present, which can be assigned to either of {the} partitions spanned by available mode functions. Consider the same prototypical example as in Eq.~\eqref{cl-ent1}:
\begin{equation}
\vert \Psi \rangle \propto 
\vert \psi_{01}(\mathbf{r}) \rangle_A \vert H \rangle_B - 
\vert \psi_{10}(\mathbf{r} ) \rangle_A \vert V \rangle_B . 
\label{cl-ent3}
\end{equation} 
For concreteness, let us assign {the} excitation to the polarization mode function, {making the available measurement} that of polarization. {Then,} the projective measurement will only be performed on the $B$-part of the {non-separable} functions, resulting in the projective measurement outcomes ${\cal M}^{H,V}_B$. The projective measurement on sub-system $A$ {is replaced} by some sort of mode selection, basis choice, or filter setting, and is a unitary operation (see {Fig.~{\ref{cl-ent}}).}

To facilitate comparison with Eq.~(\ref{q-ent}), let us further reformulate Eqs.\eqref{cl-ent1} and \eqref{cl-ent3}, highlighting the different nature of kets. The unitary operation, which {distinguishes} between two orthogonal basis states encoded via $\theta$ and $\theta + \frac{\pi}{2}$, is denoted ${\cal F}_{\theta}, {\cal F}_{\theta + \frac{\pi}{2}}$. This is a filtering operation corresponding to the choice of basis, or mode, or another unitary `sorter' operation. Assigning $0$ and $1$ to the measurement outcomes in the $\lbrace H, V\rbrace$ basis, we obtain:
\begin{equation}
\vert \Psi \rangle \propto \vert{\cal F}_{\theta} \rangle_A \vert 0 \rangle_B - \vert {\cal F}_{\theta + \frac{\pi}{2}} \rangle_A \vert 1 \rangle_B. 
\end{equation}
Comparing to Eq.~\eqref{q-ent}, we clearly see that there is no statistical correlation between two measurement outcomes here. Instead, we have a deterministic projective measurement outcome ${\cal M}^j$, given the choice of basis. There is a deterministic correlation between the `filter' setting $k=\theta, \theta + \frac{\pi}{2}$ and the measurement result, $C \lbrace k, {\cal M}^j \rbrace$.

Entanglement \textit{is} about non-separability, but the nature of non-separable functions is decisive. However, the {key point} is not whether the partitioning of the vector space, $A, B$, is related to spatially separated particles or to two degrees of freedom of the same particle. {What is important} is whether non-separable functions lead to statistically correlated measurement outcomes due to the effect of two subsequent quantum measurements, one on each sub-system (quantum entanglement, violation of realism), or {whether} mathematical non-separability leads to a single measurement outcome conditioned on some basis choice (classical entanglement, realism observed).

\section{Locality and non-locality of non-separable states}
\label{nonloc}

{In discussions comparing} classical and quantum entanglement, {distinctions are} often made using characterizations such as `local' and `non-local.' But what do they mean? This question {has sparked a controversial dispute} that is still ongoing~\cite{Khrennikov,Shen2022}. Is entanglement primarily about non-locality in the sense that an action on one part of a global system seems {to have an immediate} effect on the other part of the system? Or does 'non-local' simply mean that the system is distributed in space?

{In early 2000s, the discussion of the former aspect (is entanglement primarily about non-locality) has already led to a counter-intuitive conclusion that entanglement and non-locality are different resources. As an example, a thorough investigation of incommensurability between entanglement and non-locality is presented in \cite{Gisin}. We return to this question a bit later in this section in the context of the measurement-based approach taken.}

Regarding the latter, we have already shown in {Sec.~\ref{ent-revisited}} that the existence of two spatially separated objects is not a necessary pre-requisite for entanglement. The former is more subtle. The principle of locality states that any interaction propagates with finite velocity, and it takes time for a change in one degree of freedom to have an effect on the other degree of freedom. On the surface, any state involving mathematically non-separable functions (equally describing either spatially separated systems or the internal degrees of freedom of one and the same system) should violate this principle of locality. Therefore, the concept of ``spooky action at a distance" stands {as a characteristic feature of entanglement} from the early {Einstein-Podolsky-Rosen (EPR)}  debates in 1935 onwards. Examples of the violation of Bell-type inequalities with classically entangled light seem to confirm this \cite{Goldin2010,Khoury2010}. {This leaves anyone who attributes the term 'entanglement' to a system that is indisputably in the quantum domain quite uncomfortable. What, then, is the decisive point that } would help reliably demarcate classical correlations from genuinely quantum phenomena?

Already in this discussion of {locality and non-locality, an}  important point emerges {that} enables the demarcation. First of all, recall that the violation of the Bell inequality means that quantum mechanics is not a \textit{local realistic theory}. It is not just about locality, although we often use the term `non-locality' in connection with the Bell inequality. {This raises the question: Does} quantum mechanics violate the locality principle, the principle of reality, or {both together?}  The discussion is still ongoing, especially when different interpretations of quantum mechanics are debated. However, note: The violation of local-realistic assumptions occurs in a non-separable state only if a measurement is performed on one of the subsystems.

{As strongly emphasized in \cite{Khrennikov},} ``{the main deviation of classical light models from quantum theory is not only in the states but also in the description of measurement procedures."} {Khrennikov argues} strongly against using quantum non-locality to differentiate between quantum and classical correlations, attributing the crucial role to measurement procedures~\cite{Khrennikov}, which {aligns well with our arguments.} {This ``measurement-based'' approach provides a key to interpreting the violation of the local realism.}
Importantly, the conventional view is that quantum theory is local, {yet} the assumption of realism is incorrect. {This argument is quite straightforward}. {The} principle of locality {stipulates} that interactions propagate with finite velocity. However, when the measurement outcome on $A$ determines {the outcome} on $B$ in a non-separable states, {it actually happens not due to the interaction between these two subsystems}. Instead, a system in a non-separable state is in a superposition of two (in bi-partite systems) possibilities:  For the state of Eq.~\eqref{q-ent} the system is simultaneously in {the} state $ \vert 0 \rangle_A \vert 1 \rangle_B$ and $\vert 1 \rangle_A \vert 0 \rangle_B$, it is not in a definite, `pre-existing'  state. The distinct state emerges only upon measurement and is `decided' at random in the process of measurement with a given probability. 

{This clearly demonstrates a violation of the principle of reality, which posits that the properties of objects are inherent and exist independently of our observations. This once again underscores the importance of the quantum measurement as a fundamental aspect of the quantum-classical boundary. The manifestation of a specific `reality' during} the measurement process is the source of correlations in case of quantum entanglement. The first projective measurement, {performed } on $A$, collapses the global state in one of the possibilities, at random, say $ \vert 0 \rangle_A \vert 1 \rangle_B$, leading to random `0' at $A$. The effect of this first measurement is verified by the second measurement, which reveals the state of the second sub-system to be `1', confirming statistical correlation {(Fig.~{\ref{ent}}).}

In classical entanglement, as elucidated above, a unitary process of selecting a particular basis or mode function replaces the measurement for {certain} operations. {This process, unlike a measurement, is unitary (at least in principle), and we refer to it as ``filtering" or ``sorting". Due to the unitary nature of the operation, there is no reduction of the wave function.} Hence, the mathematical non-separability in the framework of classical theory (e.g., in classical optics involving {vector light fields}) cannot result in any non-local effects. {In general, understanding the distinction between measurement and certain unitary processes, which are sometimes confusingly interpreted as measurements, provides insight into the controversial usage of the term entanglement.}

\begin{figure}[t]
\label{Bell-fig}
\centering
\includegraphics[width=\linewidth]{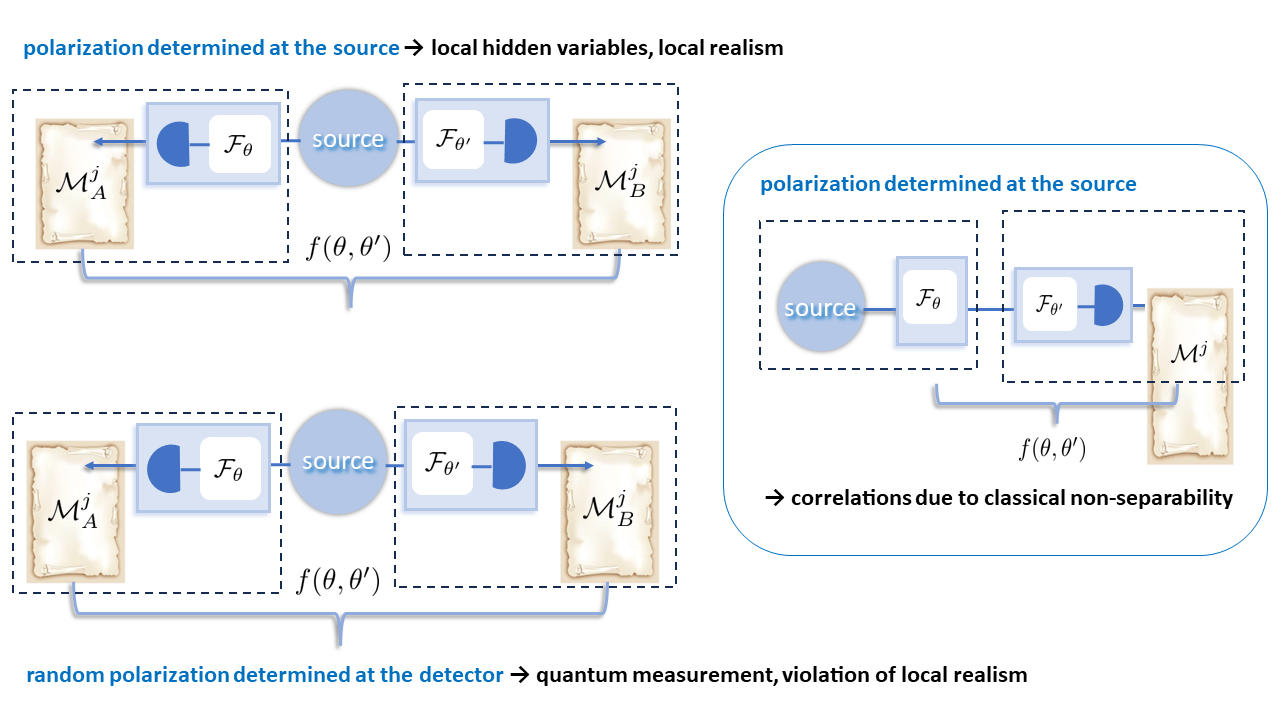}
\caption{Principle experimental schemes for construction of Bell inequality for verification of polarization correlations $f(\theta, \theta^\prime)$. Parameterization by $\theta$ happens either as part of the measurement process, or as part of the state preparation process, as indicated by dashed boxes. (Top Left) In case of LHV assumption, the polarization is predetermined at the source and two measurements are performed parameterized by $\theta$;  (Bottom  Left) In case of quantum entanglement, the polarization is not defined at the source and is determined in the measurement process, at the detector, when two measurements are performed parameterized by $\theta$; (Right) In case of classical non-separability, light at the source is prepared in a certain polarization state  parameterized by $\theta$. A single, parameterized by $\theta^\prime$ measurement is performed.}
\end{figure}

It is very tempting to exploit the analogy between wave-like properties in quantum mechanics and all the wave phenomena in classical optics (wave equation, superposition, interference, etc.). {After all, in Erwin Schrödinger's early works, quantum mechanics went} under the name wave mechanics~\cite{wellenmechanik}. This viewpoint has been widely represented in {the} community, strongly highlighted in the works of Eberly~\cite{Eberly2015,Eberly2016}.  This analogy has also been extended to demonstrations of violations of Bell inequalities. Simulations of Bell inequalities are discussed in a number of papers, for example \cite{Goldin2010,Khoury2010}. 

Let us depart for the moment from any specific experimental details {on how} it has been implemented, but let us focus on the principal idea of an experimental setting used to construct the inequality \cite{bell_aspect_2004,Paneru_2020}. The statistical analysis required {begins with} performing \textit{two} independent measurements on two parts of the Hilbert space, these measurements being parameterized by some parameter $\theta$. For example, in the case of polarization entanglement of Eq.~(\ref{singlet}), the measurement process will include a polariser, set in front of a detector and rotated at a certain angle $\theta$. {In other words, the }measurement will include a detector and a `filter' set for $\theta$, {as discussed previously.} The inequality is constructed by combining probabilities of measurement outcomes obtained at detector 1 plus filter $\theta$ and detector 2 plus filter $\theta^{\prime}$ for particular values of $\theta$ and $\theta^{\prime}$ {(for details, we refer the interested reader to \cite{bell_aspect_2004,Paneru_2020}).} 

Consider now three relevant cases of polarization correlations $f(\theta, \theta^\prime)$: determined by local hidden variables (LHV); stemming from quantum entanglement; originating from classical non-separability. The three respective schemes are depicted in {Fig.~\ref{Bell-fig}} with the example of polarization measurements. The key difference can be easily understood if we compare this figure with {Figs.~\ref{ent} and \ref{cl-ent}.} The verification of the local realism assumption has been derived by Bell {by examining} statistical correlations of the results of \textit{two} independent measurements on two partitions of the Hilbert space (left part of {Fig.~\ref{Bell-fig}).} One can simulate this test using classical non-separability, but it will not yield the same statistical analysis as performed by  {Bell; instead, it exploits deterministic} correlations between state preparation and a single measurement to emulate Bell-like correlations (right part of {Fig.~\ref{Bell-fig}).}

\section{Conclusions}

Quantum and classical optics {constitute} an amazing field for exploring the interplay between the particle and wave nature of light~\cite{Loudon}. {The richness of wave phenomena goes beyond usual applications, allowing for the emulation of quantum features, as seen in classical entanglement, and even serving as a platform for analogue gravity experiments~\cite{Koenig}. Despite the beauty of both quantum and classical theories of light,} one should not be tempted to move the quantum-classical boundary entirely on the ground of presence of wave phenomena, or existence of non-separable vector spaces, or possibility for enhancing metrology protocols.  Quantum mechanics is first and foremost about the interaction between a system and a measurement apparatus.  Not the properties of the system taken by itself, and not so much the microscopic nature of the system itself are decisive. 

{Assessing} the quantum nature of a non-separable state {requires} consideration of its behavior in a measurement process. In simple, operational language, {the} quantum nature of non-separable states {manifests} in \textit{statistical correlations} between the outcomes of \textit{two measurements} performed on two partitions of the corresponding Hilbert space. {This distinctive feature transcends the } {the details of state preparation taken by themselves} and {whether we are} dealing with single photons or intense light beams.

{The scheme} depicted at the bottom left part of {Fig.~\ref{Bell-fig}} can be utilized not only to test local realism for polarization entanglement of single photons, but also for continuous variable polarization entanglement \cite{Loundon-pol-ent}. {Such} non-separable states reveal quantum correlations in the measurement of Stokes operators, quantum variables with a continuous spectrum, the counterparts of classical Stokes parameters \cite{Loudon-nonsep}. These states can be generated using intense light beams that may appear classical.

The question arises: if classical non-separability is {entirely} a classical phenomenon and not entanglement in its true sense, then where {does} the advantage demonstrated in metrology, the possibility to simulate not only Bell-like non-locality, but also a number of other quantum information protocols, come from? This is not the subject of this paper, and we will only briefly comment on {it}. In most cases, {the} advantage comes merely from the increased parameter space. One can conceptualize it as the size and structure of the Hilbert space, {as seen in quantum cryptographic protocols based} on structured light \cite{Karimi-QKD}, or as a form of classical parallelism, as observed in polarization metrology \cite{toppel,wir2019}.

{In the same vein, this increased parameter space has been used in the pervasive field of indefinite causal order, where one can think of using one degree of freedom as control for an operation that is acting on another degree of freedom. Incorporating this control in particular tasks leads to  advantages in computation~\cite{Colnaghietal2012}, communication~\cite{Chiribellaetal2021}, and metrology~\cite{Zhaoetal2020,Goldberg:2023aa}, many of which have been experimentally realized~\cite{Rubinoetal2017,Goswamietal2020,Guoetal2020,Rubinoetal2021,Yin:2023aa}.}

We propose an operational distinction to clearly identify quantum entanglement and emphasize that two measurement are required for this identification. The projective quantum measurement processes by themselves remains a mystery: trying to find, where in the chain of events the projection happens is a moving target, and the „speed“ at which the projection happens seems to be solely determined by the response time of the detector.

\ack{{With great pleasure, we dedicate this article to Rodney Loudon.   His legacy continues to inspire scientists and to pave the way for further exploration in the realm of quantum optics.}
{Over the years, we have  {benefited} from questions, suggestions, criticism, and advice from many colleagues. Particular gratitude for help in various ways {goes} to Andrea Aiello, Gunnar Björk, Bob Boyd, Joe Eberly, Robert Fickler, Elisabeth Giacobino, Aaron Goldberg, Ebrahim Karimi, Antonio Z. Khoury, {Andrei Khrennikov}, and  Andrei  Klimov.}\\
{The work has been supported by the Scottish Universities Physics Alliance (SUPA), the Alexander von Humboldt Foundation (NK), and by the Agencia Espa\~{n}ola de Inbestigaci\'on (LLSS).}
}

\bibliographystyle{RS} 
\bibliography{refs} 

\begin{thebibliography}{99}

\bibitem{Spreeuw1997}
Spreeuw R. 1998  A classical analogy of entanglement. {\em Found. Phys.}
  \textbf{28}, 361.
(\href{http://dx.doi.org/10.1023/A:1018703709245}{10.1023/A:1018703709245})

\bibitem{Spreeuw2001}
Spreeuw R. 2001  Classical wave-optics analogy of quantum information
  processing. {\em Phys. Rev. A} \textbf{63}, 062302.
(\href{http://dx.doi.org/10.1103/PhysRevA.63.062302}{10.1103/PhysRevA.63.062302})

\bibitem{ProgressOpt}
Forbes A, Aiello A, Ndagano B. 2019  Classically Entangled Light. {\em Progress
  in Optics} \textbf{64}, 99.
(\href{http://dx.doi.org/10.1016/bs.po.2018.11.001}{10.1016/bs.po.2018.11.001})

\bibitem{AielloNJP2015}
Aiello A, et~al. 2015  Quantum-like nonseparable structures in optical beamss.
  {\em New J. Phys.} \textbf{17}, 043024.
(\href{http://dx.doi.org/10.1088/1367-2630/17/4/043024}{10.1088/1367-2630/17/4/043024})

\bibitem{wir2019}
Korolkova N, Leuchs G. 2019  Quantum correlations in separable multi-mode
  states and in classically entangled light. {\em Rep. Progr. Phys.}
  \textbf{82}, 056001.
(\href{http://dx.doi.org/10.1088/1361-6633/ab0c6b}{10.1088/1361-6633/ab0c6b})

\bibitem{Shen2022}
Shen Y, Rosales-Guzm{\'a}n C. 2022  Nonseparable States of Light: From Quantum
  to Classical. {\em Laser Photonics Rev.} \textbf{16}, 2100533.
(\href{http://dx.doi.org/10.1002/lpor.202100533}{10.1002/lpor.202100533})

\bibitem{Eberly2015}
Qian XF, Howell JC, Eberly JH. 2015  Shifting the quantum-classical boundary:
  theory and experiment for statistically classical optical fields. {\em
  Optica} \textbf{2}, 611.
(\href{http://dx.doi.org/10.1364/OPTICA.2.000611}{10.1364/OPTICA.2.000611})

\bibitem{Eberly2016}
Eberly JH, et~al. 2016  Quantum and classical optics - emerging links. {\em
  Phys. Scr.} \textbf{91}, 063003.
(\href{http://dx.doi.org/10.1088/0031-8949/91/6/063003}{10.1088/0031-8949/91/6/063003})

\bibitem{Karimi2015}
Karimi E, Boyd RW. 2015  Classical entanglement?. {\em Science} \textbf{350},
  1172--1173.
(\href{http://dx.doi.org/10.1126/science.aad7174}{10.1126/science.aad7174})

\bibitem{Berkhout2010}
Berkhout GCG, Lavery MPJ, Courtial J, Beijersbergen MW, Padgett MJ. 2010
  Efficient Sorting of Orbital Angular Momentum States of Light. {\em Phys.
  Rev. Lett.} \textbf{105}, 153601.
(\href{http://dx.doi.org/10.1103/PhysRevLett.105.153601}{10.1103/PhysRevLett.105.153601})

\bibitem{Ionicioiu2016}
Ionicioiu R. 2016  Sorting quantum systems efficiently. {\em Sci. Rep.}
  \textbf{6}, 25356.
(\href{http://dx.doi.org/10.1038/srep25356}{10.1038/srep25356})

\bibitem{Paneru_2020}
Paneru D, Cohen E, Fickler R, Boyd RW, Karimi E. 2020  Entanglement: quantum or
  classical?. {\em Rep. Prog. Phys.} \textbf{83}, 064001.
(\href{http://dx.doi.org/10.1088/1361-6633/ab85b9}{10.1088/1361-6633/ab85b9})

\bibitem{bell_aspect_2004}
Bell JS, Aspect A. 2004 {\em Speakable and Unspeakable in Quantum Mechanics:
  Collected Papers on Quantum Philosophy}.
Cambridge University Press 2nd edition.
(\href{http://dx.doi.org/10.1017/CBO9780511815676}{10.1017/CBO9780511815676})

\bibitem{Loudon}
Loudon R. 2000 {\em The Quantum Theory of Light}.
Oxford University Press 3rd edition.

\bibitem{Goldberg:2022aa}
Goldberg AZ, Grassl M, Leuchs G, S{\'a}nchez-Soto LL. 2022  Quantumness beyond
  entanglement: The case of symmetric states. {\em Phys. Rev. A} \textbf{105},
  022433--.
(\href{http://dx.doi.org/10.1103/PhysRevA.105.022433}{10.1103/PhysRevA.105.022433})

\bibitem{Zanardi1}
Zanardi P. 2001  Virtual Quantum Subsystems. {\em Phys. Rev. Lett.}
  \textbf{87}, 077901.
(\href{http://dx.doi.org/10.1103/PhysRevLett.87.077901}{10.1103/PhysRevLett.87.077901})

\bibitem{Zanardi2}
Zanardi P, Lidar DA, Lloyd S. 2004  Quantum Tensor Product Structures are
  Observable Induced. {\em Phys. Rev. Lett.} \textbf{92}, 060402.
(\href{http://dx.doi.org/10.1103/PhysRevLett.92.060402}{10.1103/PhysRevLett.92.060402})

\bibitem{Bjork}
Hessmo B, Usachev P, Heydari H, Bj{\"o}rk G. 2004  Experimental demonstration
  of single photon nonlocality. {\em Phys. Rev. Lett.} \textbf{92}, 180401.
(\href{http://dx.doi.org/10.1103/PhysRevLett.92.180401}{10.1103/PhysRevLett.92.180401})

\bibitem{getting-used}
Leuchs G. 2016  Getting used to quantum optics. {\em arXiv preprint
  arXiv:1602.03019}.
(\href{http://dx.doi.org/doi.org/10.48550/arXiv.1602.03019}{doi.org/10.48550/arXiv.1602.03019})

\bibitem{Peres1995}
Peres A. 1995  Quantum Theory: Concepts and Methods. In {\em Quantum Theory:
  Concepts and Methods} ,  p. 116. Kluwer Academic Publishers 1st edition.

\bibitem{wineland}
Monroe C, D., Meekhof, King B, Wineland DJ. 1996  A ``Schr\"odinger cat''
  superposition state of atom. {\em Science} \textbf{272}, 1131.
(\href{http://dx.doi.org/10.1126/science.272.5265.1131}{10.1126/science.272.5265.1131})

\bibitem{Khrennikov}
Khrennikov A. 2020  Quantum versus classical entanglement: eliminating the
  issue of quantum nonlocality. {\em Found. Phys.} \textbf{50}, 1762.
(\href{http://dx.doi.org/10.1007/s10701-020-00319-7}{10.1007/s10701-020-00319-7})

\bibitem{Khrennikov2}
Khrennikov A, Basieva I. 2023  Entanglement of Observables: Quantum Conditional
  Probability Approach. {\em Found. Phys.} \textbf{53}, 84.
(\href{http://dx.doi.org/10.1007/s10701-023-00725-7}{10.1007/s10701-023-00725-7})

\bibitem{Gisin}
Brunner N, Gisin N, Scarani V. 2005  Entanglement and non-locality are
  different resources. {\em New J. Phys.} \textbf{7}, 88.
(\href{http://dx.doi.org/10.1088/1367-2630/7/1/088}{10.1088/1367-2630/7/1/088})

\bibitem{Goldin2010}
Goldin MA, Francisco D, Ledesma S. 2010  Simulating Bell inequality violations
  with classical optics encoded qubits. {\em J. Opt. Soc. Am.} \textbf{27},
  779.
(\href{http://dx.doi.org/10.1364/JOSAB.27.000779}{10.1364/JOSAB.27.000779})

\bibitem{Khoury2010}
Khoury. 2010  Bell-like inequality for the spin-orbit separability of a laser
  beam. {\em Phys. Rev. A} \textbf{82}, 033833.
(\href{http://dx.doi.org/10.1103/PhysRevA.82.033833}{10.1103/PhysRevA.82.033833})

\bibitem{wellenmechanik}
Schr\"odinger E. 2020 {\em Collected Papers On Wave Mechanics}.
Montreal: Minkowski Institute Press.
(\href{http://dx.doi.org/978-1927763803}{978-1927763803})

\bibitem{Koenig}
Jacquet MJ, Weinfurtner S, K\"onig F. 2020  The next generation of analogue
  gravity experiments (Introduction to the theme issue). {\em Phil. Trans. R.
  Soc. A} \textbf{378}, 20190239.
(\href{http://dx.doi.org/10.1098/rsta.2019.0239}{10.1098/rsta.2019.0239})

\bibitem{Loundon-pol-ent}
Korolkova N, Leuchs G, Loudon R, Ralph TC, Silberhorn C. 2002  Polarisation
  Squeezing and Continuous Variable Polarization Entanglement. {\em Phys. Rev.
  A} \textbf{65}, 052306.
(\href{http://dx.doi.org/10.1103/PhysRevA.65.052306}{10.1103/PhysRevA.65.052306})

\bibitem{Loudon-nonsep}
Korolkova N, Loudon R. 2005  Nonseparability and squeezing of continuous
  polarization variables. {\em Phys. Rev. A} \textbf{71}, 032343.
(\href{http://dx.doi.org/10.1103/PhysRevA.71.032343}{10.1103/PhysRevA.71.032343})

\bibitem{Karimi-QKD}
Bouchard F, Heshami K, England D, Fickler R, Boyd RW, Englert BG,
  S{\'{a}}nchez-Soto LL, Karimi E. 2018  Experimental investigation of
  high-dimensional quantum key distribution protocols with twisted photons.
  {\em {Quantum}} \textbf{2}, 111.
(\href{http://dx.doi.org/10.22331/q-2018-12-04-111}{10.22331/q-2018-12-04-111})

\bibitem{toppel}
T\"oppel F, Aiello A, Marquardt C, Giacobino E, Leuchs G. 2014  Classical
  entanglement in polarization metrology. {\em New J. Phys.} \textbf{16},
  073019.
(\href{http://dx.doi.org/10.1088/1367-2630/16/7/073019}{10.1088/1367-2630/16/7/073019})

\bibitem{Colnaghietal2012}
Colnaghi T, D'Ariano GM, Facchini S, Perinotti P. 2012  Quantum computation
  with programmable connections between gates. {\em Phys. Lett. A}
  \textbf{376}, 2940--2943.
(\href{http://dx.doi.org/https://doi.org/10.1016/j.physleta.2012.08.028}{https://doi.org/10.1016/j.physleta.2012.08.028})

\bibitem{Chiribellaetal2021}
Chiribella G, Banik M, Bhattacharya SS, Guha T, Alimuddin M, Roy A, Saha S,
  Agrawal S, Kar G. 2021  Indefinite causal order enables perfect quantum
  communication with zero capacity channels. {\em New J. Phys.} \textbf{23},
  033039.
(\href{http://dx.doi.org/10.1088/1367-2630/abe7a0}{10.1088/1367-2630/abe7a0})

\bibitem{Zhaoetal2020}
Zhao X, Yang Y, Chiribella G. 2020  Quantum Metrology with Indefinite Causal
  Order. {\em Physical Review Letters} \textbf{124}, 190503.
(\href{http://dx.doi.org/10.1103/PhysRevLett.124.190503}{10.1103/PhysRevLett.124.190503})

\bibitem{Goldberg:2023aa}
Goldberg AZ, Heshami K, S{\'a}nchez-Soto LL. 2023  Evading noise in
  multiparameter quantum metrology with indefinite causal order. {\em Phys.
  Rev. Research} \textbf{5}, 033198.
(\href{http://dx.doi.org/10.1103/PhysRevResearch.5.033198}{10.1103/PhysRevResearch.5.033198})

\bibitem{Rubinoetal2017}
Rubino G, Rozema LA, Feix A, Ara{\'u}jo M, Zeuner JM, Procopio LM, Brukner {\v
  {C}}, Walther P. 2017  Experimental verification of an indefinite causal
  order. {\em Sci. Adv.} \textbf{3}.

\bibitem{Goswamietal2020}
Goswami K, Cao Y, Paz-Silva GA, Romero J, White AG. 2020  Increasing
  communication capacity via superposition of order. {\em Phys. Rev. Research}
  \textbf{2}, 033292.
(\href{http://dx.doi.org/10.1103/PhysRevResearch.2.033292}{10.1103/PhysRevResearch.2.033292})

\bibitem{Guoetal2020}
Guo Y, Hu XM, Hou ZB, Cao H, Cui JM, Liu BH, Huang YF, Li CF, Guo GC,
  Chiribella G. 2020  Experimental Transmission of Quantum Information Using a
  Superposition of Causal Orders. {\em Phys. Rev. Lett.} \textbf{124}, 030502.
(\href{http://dx.doi.org/10.1103/PhysRevLett.124.030502}{10.1103/PhysRevLett.124.030502})

\bibitem{Rubinoetal2021}
Rubino G, Rozema LA, Ebler D, Kristj\'ansson H, Salek S, Allard~Gu\'erin P,
  Abbott AA, Branciard C, Brukner {\v{C}}, Chiribella G, Walther P. 2021
  Experimental quantum communication enhancement by superposing trajectories.
  {\em Phys. Rev. Research} \textbf{3}, 013093.
(\href{http://dx.doi.org/10.1103/PhysRevResearch.3.013093}{10.1103/PhysRevResearch.3.013093})

\bibitem{Yin:2023aa}
Yin P, Zhao X, Yang Y, Guo Y, Zhang WH, Li GC, Han YJ, Liu BH, Xu JS,
  Chiribella G, Chen G, Li CF, Guo GC. 2023  Experimental super-{Heisenberg}
  quantum metrology with indefinite gate order. {\em Nature Physics}
  \textbf{19}, 1122--1127.
(\href{http://dx.doi.org/10.1038/s41567-023-02046-y}{10.1038/s41567-023-02046-y})

\end{thebibliography}

\end{document}